# Radio Frequency Field-Induced Enhancement of Detection Sensitivity in Silicon Nanowire Sensors

## Author Information

## Affiliations

[1] **Division of Materials Science and Engineering, Boston University, Brookline, Massachusetts 02446, USA**

[2] **Department of Physics, Boston University, 590 Commonwealth Avenue, Boston, Massachusetts 02215, USA**

[3] **FemtoDx, Inc., 8484 Wilshire Blvd, Suite 630, Beverly Hills, CA, 90211, USA**

Ang Liu,[1,3] Jingsong Shang,[2] Jiangang J. Du,[3] Shyamsunder Erramilli,[1,2,3] Pritiraj Mohanty[1,2,3]

## Contributions

Ang Liu contributed the followings: Conceptualization; Data curation (lead); Formal analysis (equal); Investigation (lead); Methodology (lead); Software; Validation (lead); Visualization (lead); Writing – original draft (lead); Writing – review & editing (equal).

Jingsong Shang contributed the followings: Conceptualization; Data curation; Formal analysis (equal); Investigation; Methodology; Software (lead); Validation; Visualization; Writing – original draft (supporting); Writing – review & editing (equal).

Jiangang J. Du contributed the followings: Conceptualization; Resources; Supervision; Writing – review & editing (equal).

Shyamsunder Erramilli contributed the followings: Conceptualization; Supervision; Writing – review & editing (equal).

Pritiraj Mohanty contributed the followings: Conceptualization (lead); Resources (lead); Supervision (lead); Writing – review & editing (equal).

Corresponding author

Correspondence to: Ang Liu, E-Mail: aliu12@bu.edu, Pritiraj Mohanty, E-Mail: mohanty@buphy.bu.edu

## Abstract

Sensitive biomarker detection in physiological fluids is often limited by Debye screening, which suppresses electrostatic signals at sensor surfaces. Here we report a sensing approach based on flexoelectric resonance in silicon nanowire field-effect transistors. An applied radiofrequency field induces strain gradients in the nanowires, generating flexoelectric polarization that is amplified at resonant frequencies. This effect enhances the sensitivity of conductance measurements to small surface charge variations associated with biomolecular binding. Using C-reactive protein as a model biomarker, we observe an order-of-magnitude improvement in detection sensitivity compared to conventional operation, with a 62% conductance increase versus 30% without radiofrequency modulation. The high-frequency field also perturbs the electrical double layer, reducing Debye screening in high-ionic-strength environments. These combined effects enable direct biomarker detection without sample dilution. This work establishes flexoelectric resonance as a general strategy for improving nanoscale biosensing performance in physiologically relevant conditions.

## Introduction

During the past few decades, point-of-care diagnostics and personalized medicine have emerged, marking significant advancements in the bioassay landscape. Due to their nanoscale dimensions, silicon nanowire field-effect transistors (SiNW FETs) can potentially address the challenges of detecting biomarkers such as C-reactive protein (CRP) at clinically relevant low concentrations.[1,2] SiNW FETs enable rapid, high-throughput, and precise diagnostics through label-free, potentiometric biosensing. These devices operate on the principle that charged biomolecules, which specifically bind to surface-immobilized antibodies on the nanowires, induce a field effect that modulates the conductance of the silicon nanowire channels. The three-dimensional geometry of the nanowires greatly enhances the surface-to-volume ratio and, consequently, the sensitivity.[3–5] One key challenge in biodetection using SiNW FETs is overcoming the Debye screening effect in high ionic strength environments such as blood or serum. Debye screening limits sensing performance by reducing the effective electrostatic influence of surface charges, as counter-ions in the electrolyte form an electrical double layer that screens the potential over the Debye length. A better understanding of, and the ability to manipulate, the electrical double layer and Debye screening at charged interfaces can significantly improve bioanalytical sensitivity.[6]

The strength of the screening effect is characterized by the Debye length ($\lambda_D$) derived from the Debye-Hückel equation:

$$\lambda_D = \sqrt{\frac{\varepsilon k_B T}{2q^2 I}} \tag{1}$$

where $\varepsilon$ is the permittivity of the aqueous medium, $k_B$ is the Boltzmann constant, $T$ is the absolute temperature in kelvin, $q$ is the elementary charge, and $I$ is the ionic strength of the electrolyte.[7] For physiological conditions such as blood samples with an ionic strength of ~150 mM, the screening length is estimated to be ~1 nm at room temperature.[8,9] For any bound protein located farther than ~1 nm from the surface of a SiNW FET sensor, its electric field and potential become effectively screened and undetectable by the sensor. Therefore, most current SiNW FET sensors rely on diluting the solution to low ionic strengths.[10–12] Researchers have attempted to mitigate Debye screening to improve detection sensitivity using surface engineering and electronic perturbation. For example, Gao et al.[13] reported reduced screening after coating the electrode surface with polyethylene glycol (PEG). A similar result was achieved by Piccinini et al.,[14] who observed enhanced sensitivity by depositing a multilayer polyelectrolyte on the sensor surface. Other researchers, such as Katz et al.,[15] utilized impedance spectroscopy instead of conventional potentiometric measurements and achieved improved detection performance in viral DNA analysis. However, although these approaches enhance detection, they often require specialized fabrication processes or complex frequency-domain analysis, limiting practical applicability.[7,16–19] A recent study by Sarker et al. employed a graphene-based sensor operating at lower frequencies (165-531 kHz) for neurotransmitter detection in artificial sweat. In contrast, our work utilizes a high-frequency RF field reaching up to 200 MHz across the silicon nanowires, thereby enhancing detection sensitivity and signal-to-noise ratio for CRP in high ionic strength environments.[20]

While mitigating Debye screening has significantly improved biosensor performance in high ionic strength environments, these approaches often involve intricate fabrication processes, sample dilution, or complex signal analysis. Sample dilution, in particular, may compromise protein three-dimensional structure. In contrast, emerging physical phenomena offer alternative pathways to enhance device capabilities, enabling more versatile and robust detection platforms. One such phenomenon is flexoelectricity, which couples mechanical strain gradients to electric polarization in nanoscale electromechanical systems. A detailed theoretical framework describing polarization and strain-gradient coupling is provided in the Supplementary Information. Unlike piezoelectricity, which occurs only in non-centrosymmetric crystals, flexoelectricity is a universal property of dielectric materials, including centrosymmetric semiconductors, as strain gradients inherently break local symmetry and induce polarization.[21]

The universality of flexoelectricity makes it particularly attractive for applications requiring precise electromechanical control. Recent studies on silicon nanowires and related dielectric nanostructures have demonstrated enhanced electromechanical responses driven by flexoelectric effects, highlighting their potential for next-generation nanoscale devices.[22] In systems where conventional piezoelectricity is absent, strain-gradient-induced polarization can be tuned to modulate electronic transport properties at the nanoscale.[23] Consequently, flexoelectricity offers a promising route to overcome the limitations of traditional piezoelectric transduction, particularly in high-frequency silicon-based systems. Both experimental and theoretical studies indicate that such polarization can significantly influence carrier dynamics and enable advanced sensing and electronic functionalities.[24]

The integration of flexoelectric effects into nanostructured and multilayered systems has enabled promising applications in energy harvesting, actuation, and sensing. For example, flexoelectric cantilever systems exhibit enhanced charge output when strain gradients are amplified through nanoscale engineering, demonstrating improved sensitivity and robustness compared to traditional piezoelectric counterparts.[25] Moreover, the application of flexoelectric materials has expanded into diverse fields, ranging from nanoelectronics to biomedical devices, driven by their functional versatility and advancements in heterogeneous and composite structures. These developments have broadened the capabilities of flexoelectric materials and highlighted their critical role in addressing the challenges of multifunctionality and miniaturization in modern nanodevice design.[26]

This work demonstrates a sensing approach using SiNW FET sensors for biomarker detection in high ionic strength environments. We investigate the resonant behavior of the SiNW FET sensor and exploit resonant frequencies for CRP detection. Our results show that this frequency-domain method outperforms the conventional potentiometric measurements, owing to direct modulation of Debye screening by the external RF electric field and flexoelectric-induced polarization. These findings align with the work of Fraikin et al.,[27] which advances the understanding of electrochemical interfaces in high ionic environments. The selectivity of this approach is further validated using bovine serum albumin (BSA) at the same concentration, with no comparable response observed. This frequency-domain sensing strategy offers strong potential for direct biomarker detection in physiological solutions.

## Results

### Device characterization and RF response

The characterization and operational schematic of the SiNW FET for biomarker detection are illustrated in Fig. 1. Fig. 1(a) displays an optical image of a set of 24 parallel SiNW FET arrays, highlighting the source (S), drain (D), side gate terminals for applying external radiofrequency (RF) electric field modulation, and the ground (GND) terminal. A magnified top view obtained using a scanning electron microscope (SEM) shows three uniform nanowires with a width of approximately 55 nm. The local side gate electrodes, through which an external oscillating RF electric field is applied, are also used to tune the conductance of the SiNW FET by modulating charge carrier accumulation and depletion within the channel. In Fig. 1(b), the structural characterization of the SiNW FET is shown in a cross-sectional view using a false-colored SEM micrograph prepared via focused ion beam (FIB) milling. The inset presents a single nanowire covered by a 5 nm-thick layer of ALD-deposited $Al_2O_3$.[28]

In Fig. 1(c), the cross-sections of the nanowires are analyzed using transmission electron microscopy (TEM), with the sample lamella prepared by focused ion beam (FIB) milling. The internal structure, morphology, and dimensions of the nanowires are characterized from these TEM images. A scanning transmission electron microscopy high-angle annular dark-field (STEM-HAADF) image resolves atomic lattice features, indicating the high crystallinity of the nanowires. A lattice spacing of 3.1 Å between adjacent planes is consistent with the silicon [110] orientation. An inset high-resolution TEM (HRTEM) image shows the cross-sectional profile of a silicon nanowire. The nanowire dimensions are measured to be approximately 40 nm in height and 20 nm in width, corresponding to a cross-sectional area of ~600 $nm^2$. Additional details of the TEM specimen preparation using focused ion beam (FIB) techniques are provided in the Supplementary Information (Fig. S1). Fig. 1(d) presents a schematic representation of the effect of Debye screening on charged antigen biomarkers, which influences potentiometric measurements in the SiNW FET. In the absence of external perturbations, electrolyte ions accumulate near oppositely charged biomolecules, forming an electrical double layer (EDL) and leading to electrostatic screening. The surface potential decays exponentially over the Debye length. Fig. 1(e) depicts a schematic of the RF measurement configuration, represented as an equivalent circuit model with dual side gates adjacent to the nanowires for modulation via external RF fields.

**Fig. 1: Characterization and operational schematic of the SiNW FET for biomarker detection.**

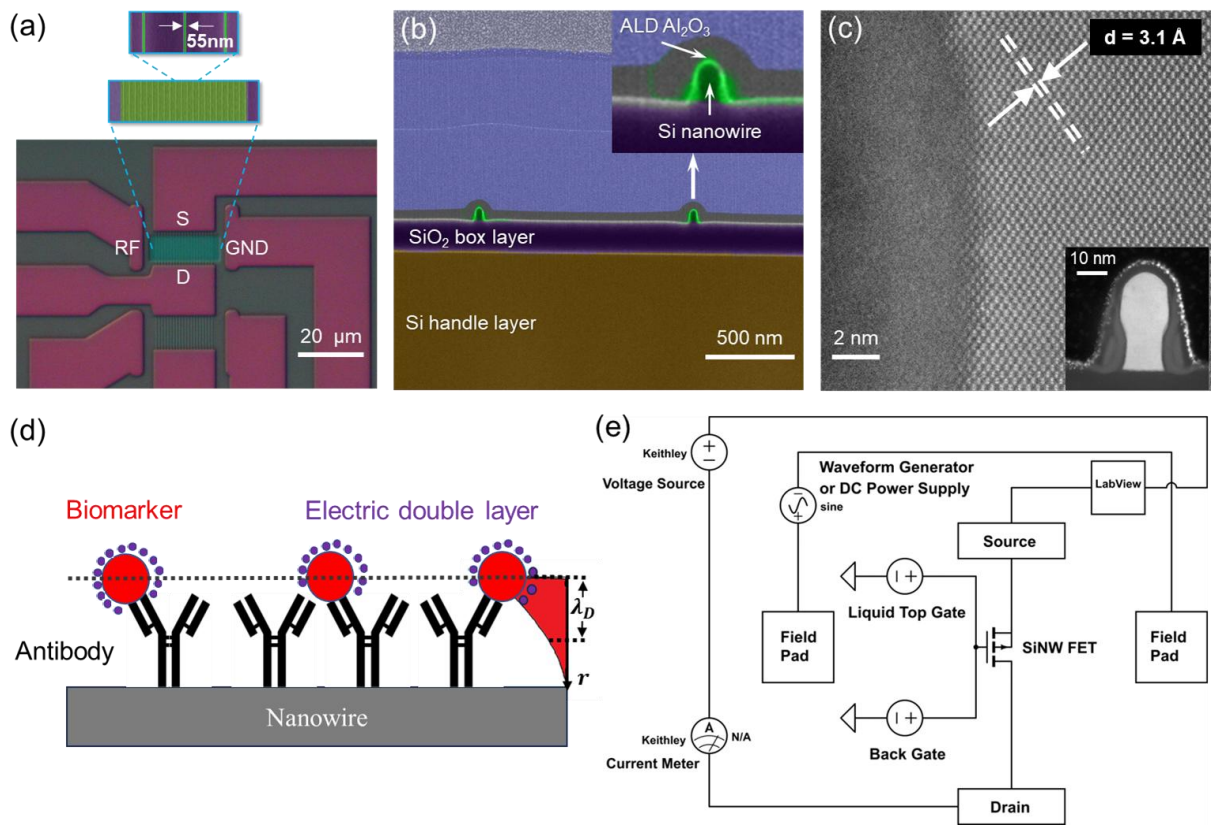

(**a**) Optical image of a fabricated SiNW FET, highlighting the source (S), drain (D), and side gate terminals (RF, GND), with a magnified SEM view of three nanowires. (**b**) False-colored SEM micrograph showing a cross-sectional view of the SiNW FET, with an inset of a single nanowire covered by a 5 nm-thick $Al_2O_3$ layer. (**c**) STEM HAADF image resolving atomic lattice features, indicating high crystallinity of the nanowires. A lattice spacing of 3.1 Å between adjacent planes is consistent with the silicon [110] orientation. The inset shows an HRTEM image of the nanowire cross-sectional profile. (**d**) Schematic representation of the effect of Debye screening on charged antigen biomarkers and its influence on potentiometric measurements in the SiNW FET. (**e**) Schematic overview of the RF measurement configuration, featuring dual side gates adjacent to the nanowires for modulation via external RF fields.

## RF-induced resonance and enhanced biomarker detection

Fig. 2(a) shows the device conductance under various conditions as the RF frequency is swept from 0 to 50 MHz. A pronounced resonance is observed, with a main peak at 10.5 MHz and several troughs around 36.8 MHz. A detailed quantitative analysis of peak positions, widths, and amplitudes for CRP and control conditions is provided in the Supplementary Information (Table S1). The peak conductance significantly increases upon application of CRP solution (~59 μS), compared to dry and wet control (1x PBS) conditions (~17 μS), indicating a strong response to CRP presence. A power-dependent study in Fig. 2(b), conducted over a broad frequency range, reveals additional peaks and troughs near 60 and 125 MHz. Fig. 2(c) shows the conductance of silicon nanowires under varying RF power levels, highlighting the behavior at 10.5 MHz. This frequency is selected for RF-modulated CRP detection due to its strong resonance peak. Figs. 2(d) and 2(e) present time-series data for CRP detection with and without RF modulation, using concentrations ranging from 0.1 pM to 1 μM. Fig. 2(f) compares the two measurement approaches. At low CRP concentrations, the conductance values are nearly identical in both cases (0.55% difference), indicating that the RF field does not directly alter the charge carrier density in the SiNW FET. However, above 0.1 nM CRP, the conductance with RF modulation increases rapidly, while the response without RF remains low. This enhancement in biomarker detection is attributed to RF-induced modulation, which disrupts the rapid reformation of the electrical double layer (EDL) surrounding the target biomolecules, consistent with prior reports.[7]

Fig. 2(b) demonstrates the RF power dependence of conductance, providing insight into high-frequency RF effects on nanowire biosensing. Experiments comparing device conductance at different CRP concentrations, with and without the external RF field, highlight significant differences in conductance between adjacent concentrations at resonant peaks or characteristic frequencies, as shown in Figs. 2(d)-2(f). These results indicate that, within clinically relevant CRP concentration ranges, operation at a characteristic RF frequency yields improved sensitivity compared to measurements without an RF electric field. The extracted resonance parameters and their variation across conditions are summarized in Table S1 of the Supplementary Information.

**Fig. 2: C-reactive protein (CRP) detection using RF-modulated SiNW FET.**

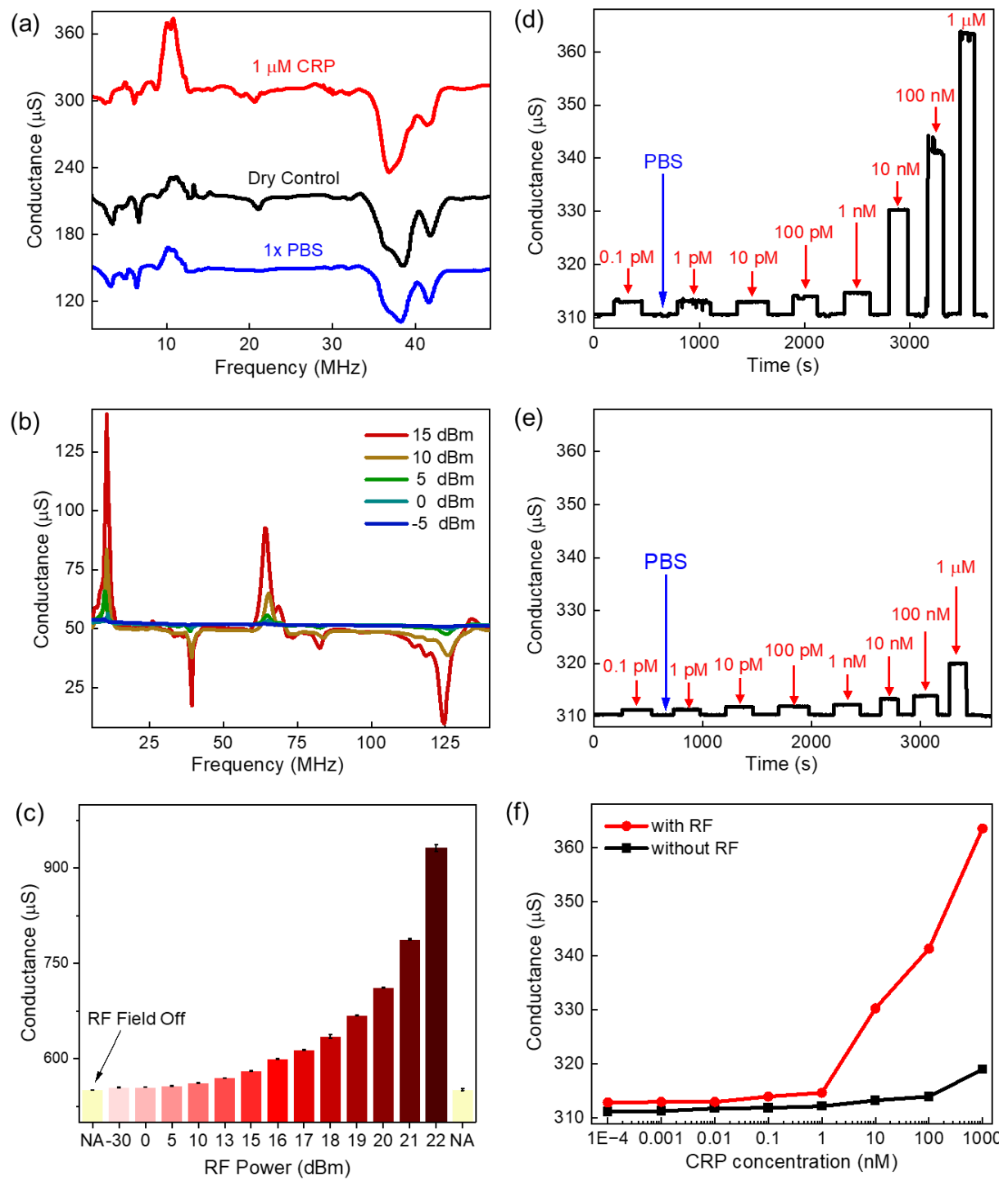

(**a**) Device conductance vs. RF frequency under various conditions (1 μM CRP, dry condition, and 1x PBS), measured using a Keithley 2400 SourceMeter, showing peaks at 10.5 MHz and troughs at 36.8 MHz. The drain-source voltage ($V_{DS}$) is 0.7 V, and the gate-source voltage ($V_{GS}$) is 0.8 V. (**b**) Device conductance vs. RF frequency at power levels from -5 to 15 dBm. (**c**) Bar graph showing the dependence of conductance of the silicon nanowires on RF power at 10.5 MHz. (**d**) Time-series data of CRP detection using RF-modulated SiNW FET, where the RF field set to 10.5 MHz at 0 dBm. (**e**) Time-series data of CRP detection using conventional SiNW FET measurements, where the RF field turned off. (**f**) Comparison between RF-modulated and conventional DC measurements, showing enhanced detection of CRP at concentrations above 1 nM. Error bars are omitted for clarity, as they are smaller than the symbol size.

## Sensitivity and selectivity of detection

Further investigation into the sensitivity and selectivity of this RF-modulated sensing method is presented in Fig. 3. Fig. 3(a) shows the device conductance during CRP detection at higher frequencies ranging from 40-90 MHz. As the CRP concentration increases from 0.5 nM to 1 μM, the device conductance rises by 62.5%. Peak analysis is performed on the resonance near 65 MHz, and the peak area, along with the baseline, is plotted across different CRP concentrations in Fig. 3(b). A positive correlation between peak area and CRP concentration is observed, which is attributed to enhanced resonant effects from increased protein binding on the surface. Peak area analysis provides a frequency-independent approach for CRP detection, exhibiting a concentration dependence consistent with baseline conductance.

Moreover, to demonstrate selective detection of CRP, experiments were conducted in which BSA solutions of varying concentrations were introduced into the fluidic chamber atop the device. The resulting conductance measurements are shown in Fig. 3(c). Compared to the frequency response of CRP in Fig. 3(a), the BSA response differs markedly in both peak position and peak area. The observed peaks in the CRP and BSA spectra arise from multiple mechanisms influencing conductance, including variations in the local dielectric environment, charge carrier oscillations in the nanowires, and changes in carrier mobility and scattering. Further work is required to isolate and quantify these contributions. Fig. 3(d) presents the device conductance as a function of protein concentration for both BSA and CRP solutions. As the BSA concentration increases, the conductance of the nanowires rises from $82.55 \pm 1.52$ μS to $107.76 \pm 1.59$ μS, likely due to nonspecific binding. In contrast, for CRP, the conductance increases from 96.42 ± 1.05 μS to 156.08 ± 2.53 μS. This pronounced difference arises from specific binding between CRP and the functionalized channel surface, resulting in significantly greater response compared to nonspecific interactions. These results demonstrate the selectivity of the RF-modulated detection method using SiNW FETs.

**Fig. 3: Sensitivity and selectivity study of the RF-modulated SiNW FET biosensing system.**

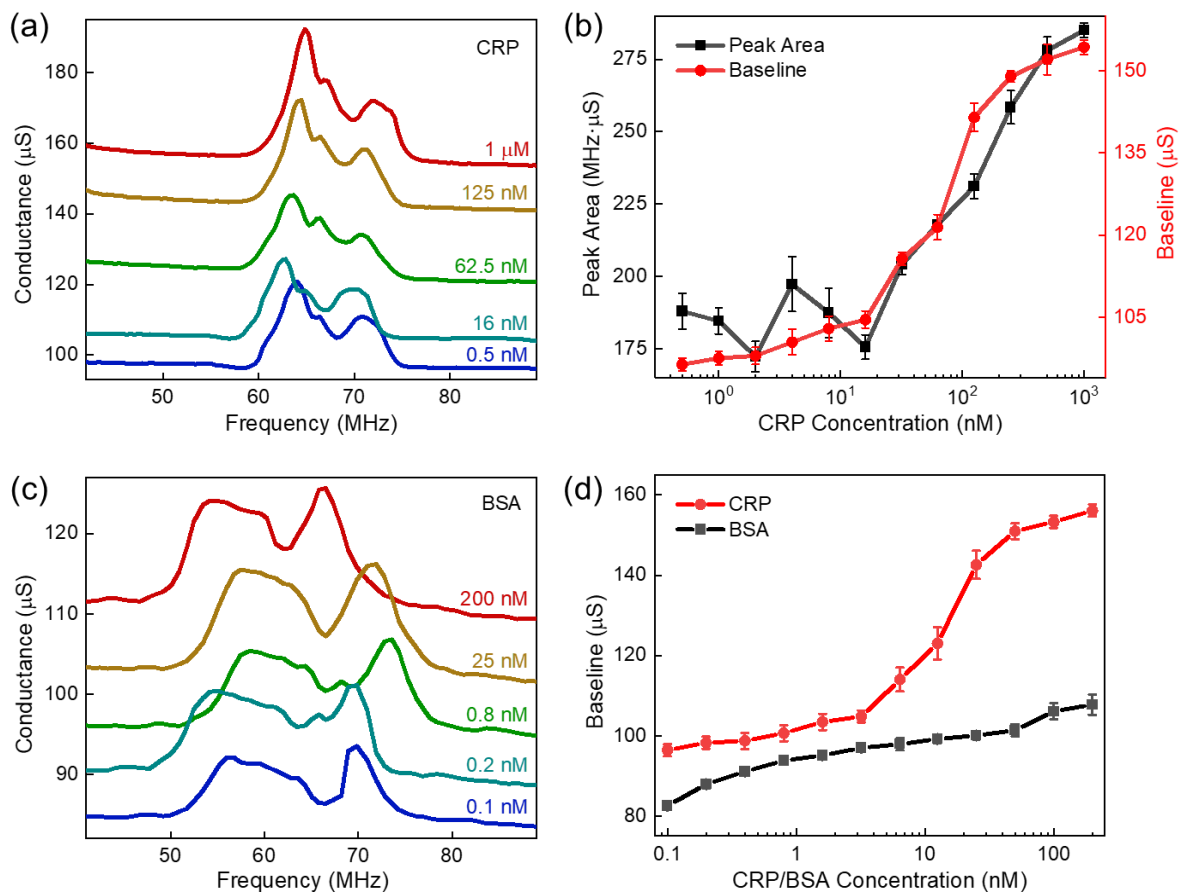

(**a**) Device conductance vs. RF frequency for a CRP dilution series (0.5-1000 nM), measured using a Keithley 2400 SourceMeter. (**b**) Baseline and peak area analyses showing comparable effectiveness for CRP detection. Connecting lines are drawn to guide the eye. (**c**) Device conductance vs. RF frequency for a BSA dilution series (0.1-200 nM), measured using a Keithley 2400 SourceMeter. (**d**) RF-modulated SiNW FET sensor response for CRP and BSA detection. The change of baseline conductance in the CRP case indicates specific binding, compared to the slight increase in the BSA case due to non-specific binding.

## Flexoelectric mechanism and resonant modes

To better understand the mechanisms behind the enhanced biosensing performance of SiNW FETs, we fabricated a device to visualize and quantify flexoelectric effects under an applied RF field. Fig. 4 illustrates how mechanical strain gradients in the centrosymmetric silicon lattice generate an internal flexoelectric polarization field, highlighting its role in coupling mechanical stress with electronic transport. In Fig. 4(a), schematics illustrate the flexoelectric polarization field induced by strain gradients within the centrosymmetric silicon crystal lattice. This field arises from flexoelectricity, where mechanical-electrical coupling becomes significant at the nanoscale. The resulting polarization provides a means to modulate electronic transport properties even in centrosymmetric materials. The diagram emphasizes how flexoelectricity bridges mechanical strain and electronic behavior, enabling enhanced tunability of SiNW-based biosensors.

The crystal lattice is shown in Fig. 4(b), where strain, structural modifications, and electronic polarization collectively influence device behavior. The flexoelectric polarization induced by the applied RF field arises from strain-induced lattice distortions, as illustrated in the schematic and supported by experimental observations. The resulting polarization modulates electronic transport, as evidenced by the conductance shifts observed in Figs. 2 and 3. Unlike piezoelectric effects, flexoelectricity operates in centrosymmetric semiconductors such as silicon, enabling tuning of electrical properties across a broader class of materials. The flexoelectric behavior demonstrated in these nanoscale devices, specifically SiNW biosensors, highlights strong potential for next-generation, strain-sensitive nanoscale systems, particularly in centrosymmetric materials.

Fig. 4(c) shows distinct harmonic conductance peaks corresponding to the fundamental and higher-order resonant modes as the RF frequency is swept across the silicon nanowires. Flexoelectric polarization is expected to be enhanced at resonant frequencies, and the observed peaks are consistent with this theoretical prediction. Fig. 4(d) presents a comparison between measured and predicted resonant frequencies for different flexural modes, demonstrating an excellent quadratic relationship on a linear-linear scale. When plotted against $(2n + 1)^2$, a near-linear fit with a slope of 0.3911 is obtained, consistent with theoretical predictions based on a base frequency of 3.52 MHz. A full comparison of measured and theoretical resonance frequencies across multiple flexural modes is provided in the Supplementary Information (Table S2). This agreement further substantiates the critical role of flexoelectricity, particularly flexural modes, in governing nanoscale electromechanical coupling, providing key insights for the design of ultrahigh-sensitivity RF-modulated silicon nanobiosensors.

**Fig. 4: Flexoelectric mechanism and flexural mode analysis.**

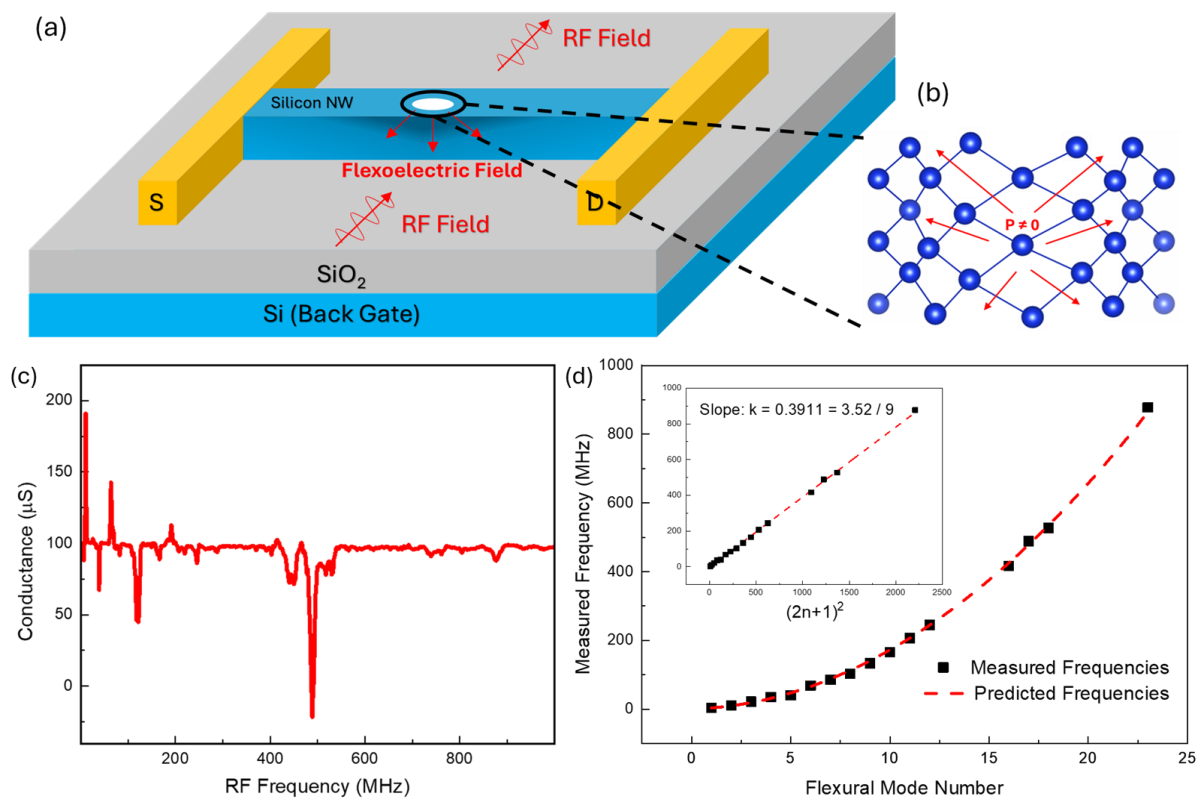

(**a**) Schematics illustrate flexoelectric effects in silicon nanowires (SiNW), with a side view showing the flexoelectric polarization field under an externally applied RF field. (**b**) Effects of flexoelectricity on crystal structure and electronic transport, illustrating strain-induced lattice distortions and the flexoelectric polarization in centrosymmetric crystals, which modulates electronic transport. (**c**) RF frequency sweep showing the conductance response, with harmonic peaks indicating resonant frequencies. (**d**) Flexural mode analysis with a base frequency of 3.52 MHz, comparing measured frequencies (data points) and predicted frequencies (dashed line) across mode numbers, demonstrating an excellent fit.

## Simulated strain gradients and polarization

To corroborate the frequency-dependent behavior observed in our SiNW FET measurements, we performed COMSOL simulations to visualize the evolution of strain gradients and the resulting flexoelectric polarization under resonant RF conditions. The material parameters and simulation inputs used in the COMSOL modelling are provided in the Supplementary Information (Table S4). These simulations confirm the critical role of flexural modes in enhancing electromechanical coupling. In particular, Fig. 5(a)-5(c) illustrate distinct flexural mode shapes in silicon nanowires (not to scale), each corresponding to a specific resonant frequency. These vibrational patterns provide insight into the dynamic electromechanical behavior of the structure under RF modulation. The flexural modes, obtained through finite element analysis (FEA) modelling, capture the strain gradient distributions that drive flexoelectric polarization. In this context, flexoelectricity links the simulated mode shapes with the experimentally observed resonant frequencies, reinforcing the underlying mechanism. Overall, the simulation results validate the flexoelectricity-driven behavior in nanoscale structures and highlight their central role in understanding electromechanical coupling mechanisms.

Additional simulation outputs are shown in Figs. 5(d) and 5(e), illustrating the strain distribution and the corresponding flexoelectric polarization across the silicon nanowires, respectively. In Fig. 5(d), localized regions of high strain gradients are observed near the binding sites on the nanowire surface, consistent with the flexural mode shapes shown in Fig. 5(a)-5(c). These strain gradients give rise to flexoelectric polarization, as illustrated in Fig. 5(e). The resulting polarization profile closely follows the strain gradient distribution and provides a quantitative measure of the electromechanical coupling strength. These results indicate that strain gradients enable precise modulation of electronic transport, leading to distinct flexoelectric polarization behavior in silicon nanowires and highlighting its critical role in achieving high-sensitivity device performance. The visualized strain and polarization distributions provide direct insight into flexoelectricity in SiNW systems and its nanoscale electromechanical effects. This insight supports the development of next-generation RF-modulated SiNW biosensors with enhanced sensitivity and selectivity.

**Fig. 5: COMSOL simulations illustrate flexural mode shapes and flexoelectric effects.**

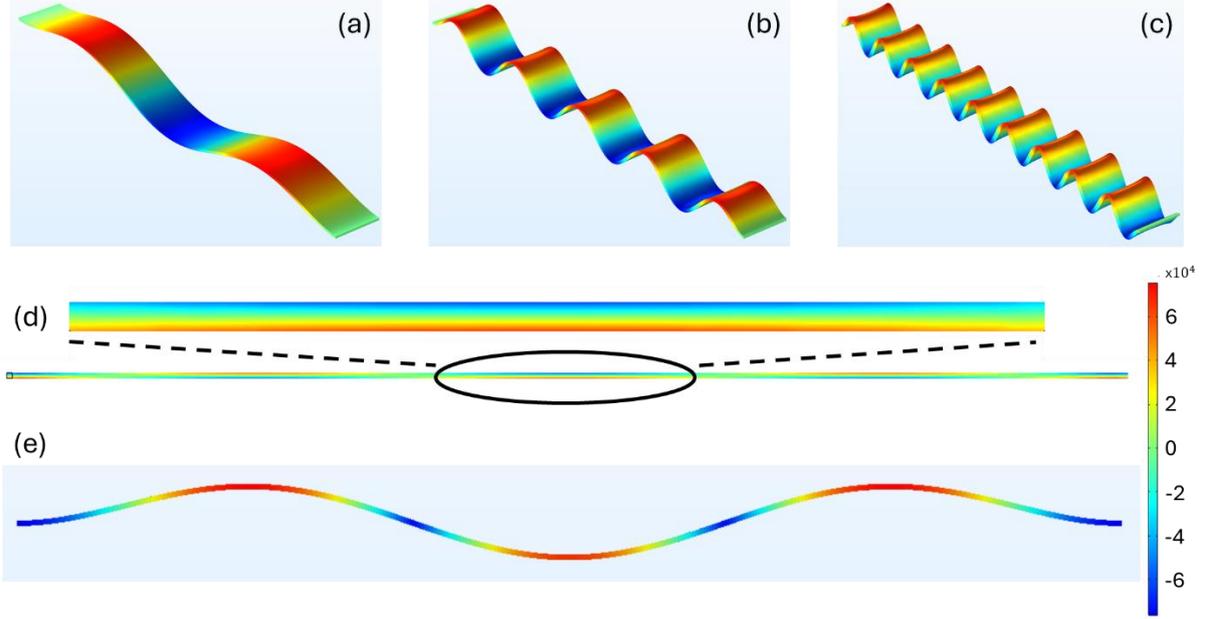

(**a-c**) Simulated flexural mode shapes at increasing resonant frequencies, with higher-order modes appearing at progressively higher frequencies. These mode shapes provide insight into the resonance characteristics and vibrational behavior of the silicon nanowires. (**d**) Strain distribution across silicon nanowires, showing localized regions of high strain gradients. (**e**) Corresponding flexoelectric polarization distribution, highlighting electromechanical coupling driven by the strain gradients.

## Discussion

To understand the underlying mechanism of Debye screening modulation, we analyze ion transport under flow and external fields in the fluidic chamber using the coupled Poisson-Nernst-Planck (PNP) equations.[29]

$$\frac{\partial \rho_i(\mathbf{r},t)}{\partial t} = \nabla \cdot [D_i(\mathbf{r})(\nabla \rho_i + \beta q_i \rho_i \nabla \varphi)] \quad (2)$$

where $\rho_i(\mathbf{r},t)$ denotes the concentration field of ionic species $i$, $D_i(\mathbf{r})$ is the corresponding diffusion coefficient, $\beta = 1/(k_B T)$, and $\varphi(\mathbf{r})$ is the electric potential. The full formulation of the governing equations, including polarization, strain, and strain-gradient relations, is provided in the Supplementary Information (Equations S1-S5). These equations describe the coupled processes of ionic diffusion and migration under electrostatic forces. The diffusion coefficient is related to ionic mobility through the Stokes-Einstein relation, where mobility depends on the hydrodynamic size and shape of the ions. Larger or more strongly hydrated species exhibit lower mobility and slower transport dynamics. The characteristic Debye relaxation time, which describes the timescale over which ionic distribution returns to equilibrium, determines the system's response to external electric fields. When the frequency of the applied RF field approaches or exceeds this relaxation timescale, the formation and reorganization of the electrical double layer (EDL) are disrupted, leading to modulation of Debye screening.

At high frequencies, i.e., in the radio frequency (RF) regime, the electric field applied across the lateral electrodes is proposed to generate non-equilibrium (high-energy) electron populations. Due to their significantly lower mass, electrons can respond to high-frequency (MHz-range) oscillations, whereas the heavier ionic species remain largely immobile on these timescales. These energetic electrons can perturb ionic interactions within the electrical double layer, contributing to partial disruption of the Debye screening layer (Stern layer and diffusion layer) and modifying local electrostatic conditions near the sensor surface. Our results indicate that, under high-frequency excitation, Debye screening is effectively mitigated, enhancing the sensitivity of conductance signals to changes in surface charge density. Regarding thermal effects, although the effective electron temperature may be elevated, the total electron energy remains limited due to their small mass and low number density relative to the bulk ionic solution. Consequently, energy transfer to the surrounding fluid is minimal, and macroscopic heating is expected to be negligible. These RF-induced effects offer significant advantages for high-frequency operation of SiNW FET biosensors. Combined with the ability to resolve small conductance variations and advances in semiconductor fabrication, this approach enables promising opportunities for nanowire-based biosensors operating in both DC and high-frequency AC regimes, even in physiological ionic environments.

Another possible contribution arises from flexoelectric polarization, which originates from strain gradients independent of crystal symmetry and can act as an effective gating mechanism in silicon nanowire FETs. The mathematical description of flexoelectric polarization and its dependence on strain gradients is detailed in the Supplementary Information. Near resonance, this effect enhances sensitivity to surface charge variations, such as those associated with protein biomarkers. In this work, we implement an all on-chip design in which side-gate electrostatic actuation induces mechanical strain, generating polarization charges that increase channel conductance.[22] This approach serves as a scalable alternative to atomic force microscope (AFM) probe indentation for creating local strain gradients.[30] A DC field applied between the side gate and the nanowire channel (Fig. 1(a)) induces strain along the nanowire, leading to flexoelectric polarization primarily in the gate oxide layer ($Al_2O_3$), and to a lesser extent in the silicon. This polarization modulates the local electrostatic environment, effectively gating the channel and increasing conductance (Fig. 2). The gating effect is further amplified by the nanoscale beam geometry, which enhances surface interactions. An RF field applied between the side gate and the nanowire induces flexoelectric polarization that follows the oscillating field. At resonance, this effect is further enhanced by the mechanical quality factor (Q). The flexoelectric coefficients and material parameters used for modeling are summarized in the Supplementary Information (Table S3 and S4).

The critical role of flexoelectricity in amplifying device sensitivity is demonstrated through the experimental and simulation results of RF-modulated resonant modes. At resonant frequencies, strain gradients are maximized, leading to enhanced flexoelectric polarization. As shown in Figs. 2, 3, and 5, the flexoelectricity-enhanced effects are reflected in the sharp conductance peaks and troughs observed in the RF frequency sweeps. Biomarker detection at extremely low concentrations is optimized by aligning the external RF field with the resonant flexoelectric modes, thereby strengthening electromechanical coupling. These findings highlight the potential of leveraging intrinsic flexoelectric properties in nanoscale structures to design next generation nanobiosensors, where resonant behavior serves as a fundamental sensing mechanism.

The interrelationship between strain gradients, flexoelectricity, and electronic transport has significant implications beyond biosensing. Potential applications include precision actuators, energy harvesting, and other advanced electromechanical systems that leverage flexoelectric effects. The strain gradient and flexoelectric polarization fields provide a framework for exploring related mechanisms in other centrosymmetric materials. Future studies may focus on quantifying flexoelectric effects on electronic transport under varying environmental conditions and examining the influence of crystal lattice modifications on flexoelectric behavior. These insights pave the way for the design and fabrication of highly sensitive, multifunctional nanodevices.

In conclusion, we demonstrate a frequency-domain method that harnesses flexoelectric effects to significantly enhance the sensitivity and selectivity of silicon nanowire field-effect transistor (SiNW FET) sensors for biomarker detection. By applying an external RF field at resonant frequencies, we modulate the Debye screening in high-ionic-strength solutions, enabling sensitive detection of C-reactive protein (CRP) across multiple concentrations. This improvement is evident in both baseline conductance and the peak area of the FET frequency spectrum and remains robust when BSA is used as a control, indicating minimal nonspecific binding. The strain-induced flexoelectric polarization in centrosymmetric silicon nanowires emerges as the key mechanism, providing a scalable and universal means of modulating electronic transport. Our combined experimental and computational results establish a new pathway for designing next-generation electromechanical biosensors that leverage flexoelectric properties, enabling direct biomarker detection in complex physiological media such as blood or serum, and offering broad potential for biomedical and industrial applications. Additional experimental details, theoretical derivations, and quantitative analyses are provided in the Supplementary Information.

## Methods

### Semiconductor device fabrication

The silicon nanowire devices used in our experiments are fabricated using complementary metal-oxide-semiconductor (CMOS)-compatible top-down semiconductor processes. The devices are fabricated on an 8-inch (100) silicon-on-insulator (SOI) wafer (Soitec, Inc.). The wafer consists of a single-crystalline silicon device layer with a thickness of 100 nm and a volume resistivity of 14-22 $\Omega\cdot$cm, a 200 nm buried oxide (BOX) layer, and a 700 μm-thick boron-doped silicon handle substrate.

The fabrication steps for the silicon nanowire devices, with an average wire width of 50-60 nm, are as follows. First, the electrodes, side gates, and nanowire regions are defined using photolithography and electron-beam lithography masks. This is followed by deposition of an etch mask via physical vapor deposition, and pattern transfer using anisotropic reactive ion etching (RIE). Finally, a 5 nm-thick $Al_2O_3$ film is deposited over the device as the top-gate dielectric using atomic layer deposition (ALD), which suppresses gate leakage current and improves dielectric isolation between the silicon nanowires and the biofluidic environment.

Surface functionalization

The silicon nanowire surface is functionalized via hydroxylation and silanization, followed by antibody immobilization to enable specific detection of the charged protein biomarker CRP in phosphate-buffered saline (PBS). Hydroxylation is performed using RCA-1 solution to introduce hydroxyl groups onto the ALD-deposited $Al_2O_3$ surface. Silanization is carried out by immersing the chip in a 2% (v/v) (3-aminopropyl)triethoxysilane (APTES) solution in anhydrous toluene within an inert-gas-filled glovebox for one hour. The device is then rinsed multiple times with anhydrous toluene and isopropanol (IPA) to remove residual APTES, dried with nitrogen gas, and baked at 110 °C for two hours. After silanization, the surface is functionalized with primary amine groups. For antibody immobilization, anti-CRP antibodies are conjugated to the device surface via 1-ethyl-3-(3-dimethylaminopropyl)carbodiimide (EDC)/N-hydroxysulfosuccinimide (sulfo-NHS) zero-length crosslinking. A droplet of anti-CRP solution (2.5 mg/mL) is applied to the sensing area and incubated to facilitate binding of antibody molecules. Surface morphology and potential evolution during functionalization were further characterized using AFM and Kelvin probe force microscopy (KPFM), as shown in Supplementary Information (Fig. S2). Following surface functionalization, a polydimethylsiloxane (PDMS) flow chamber (~ 8 μL) is cast from a 3D-printed mold and mounted onto the device. The inlet and outlet are created using a biopsy punch to form microfluidic channels. A syringe pump is used to perform continuous flow experiments over the functionalized nanowires within the fluid chamber.

## Data Availability

The data supporting the findings of this study are available from the corresponding authors upon reasonable request.

## References

1 C. Maedler, D. Kim, R.A. Spanjaard, M. Hong, S. Erramilli, and P. Mohanty, "Sensing of the Melanoma Biomarker TROY Using Silicon Nanowire Field-Effect Transistors," ACS Sens. 1(6), 696–701 (2016).

2 Y. Chen, X. Wang, S. Erramilli, P. Mohanty, and A. Kalinowski, "Silicon-based nanoelectronic field-effect pH sensor with local gate control," Applied Physics Letters 89(22), 223512 (2006).

3 Y. Chen, X. Wang, M.K. Hong, S. Erramilli, P. Mohanty, and C. Rosenberg, "Nanoscale field effect transistor for biomolecular signal amplification," Applied Physics Letters 91(24), 243511 (2007).

4 G. Zheng, F. Patolsky, Y. Cui, W.U. Wang, and C.M. Lieber, "Multiplexed electrical detection of cancer markers with nanowire sensor arrays," Nat Biotechnol 23(10), 1294–1301 (2005).

5 K.A. Muratore, D. Zhou, J.J. Du, J.S. Chlystek, K. Motesadi, E.K. Larsen, B.M. Molgora, T.C. Lee, S. Pamarti, S. Erramilli, and P. Mohanty, "Alanine aminotransferase assay biosensor platform using silicon nanowire field effect transistors," Commun Eng 2(1), 1–10 (2023).

6 D.C. Grahame, "The Electrical Double Layer and the Theory of Electrocapillarity.," Chem. Rev. 41(3), 441–501 (1947).

7 G.S. Kulkarni, and Z. Zhong, "Detection beyond the Debye Screening Length in a High-Frequency Nanoelectronic Biosensor," Nano Lett. 12(2), 719–723 (2012).

8 B. Jagannath, S. Muthukumar, and S. Prasad, "Electrical double layer modulation of hybrid room temperature ionic liquid/aqueous buffer interface for enhanced sweat based biosensing," Analytica Chimica Acta 1016, 29–39 (2018).

9 K. Maehashi, Y. Ohno, and K. Matsumoto, "Utilizing research into electrical double layers as a basis for the development of label-free biosensors based on nanomaterial transistors," Nanobiosensors in Disease Diagnosis 5, 1–13 (2015).

10 H. Chen, X. Zhao, Z. Xi, Y. Zhang, H. Li, Z. Li, H. Shi, L. Huang, R. Shen, J. Tao, and T. Wang, "A new biosensor detection system to overcome the Debye screening effect: dialysis-silicon nanowire field effect transistor," International Journal of Nanomedicine 14, 2985–2993 (2019).

11 N. Lloret, R.S. Frederiksen, T.C. Møller, N.I. Rieben, S. Upadhyay, L.D. Vico, J.H. Jensen, J. Nygård, and K.L. Martinez, "Effects of buffer composition and dilution on nanowire field-effect biosensors," Nanotechnology 24(3), 035501 (2012).

12 Y.-W. Huang, C.-S. Wu, C.-K. Chuang, S.-T. Pang, T.-M. Pan, Y.-S. Yang, and F.-H. Ko, "Real-Time and Label-Free Detection of the Prostate-Specific Antigen in Human Serum by a Polycrystalline Silicon Nanowire Field-Effect Transistor Biosensor," Anal. Chem. 85(16), 7912–7918 (2013).

13 N. Gao, T. Gao, X. Yang, X. Dai, W. Zhou, A. Zhang, and C.M. Lieber, "Specific detection of biomolecules in physiological solutions using graphene transistor biosensors," Proceedings of the National Academy of Sciences 113(51), 14633–14638 (2016).

14 E. Piccinini, S. Alberti, G.S. Longo, T. Berninger, J. Breu, J. Dostalek, O. Azzaroni, and W. Knoll, "Pushing the Boundaries of Interfacial Sensitivity in Graphene FET Sensors: Polyelectrolyte Multilayers Strongly Increase the Debye Screening Length," J. Phys. Chem. C 122(18), 10181–10188 (2018).

15 E. Katz, and I. Willner, "Probing Biomolecular Interactions at Conductive and Semiconductive Surfaces by Impedance Spectroscopy: Routes to Impedimetric Immunosensors, DNA-Sensors, and Enzyme Biosensors," Electroanalysis 15(11), 913–947 (2003).

16 G. Zheng, X.P.A. Gao, and C.M. Lieber, "Frequency Domain Detection of Biomolecules Using Silicon Nanowire Biosensors," Nano Lett. 10(8), 3179–3183 (2010).

17 D.P. Tran, M. Winter, C.-T. Yang, R. Stockmann, A. Offenhäusser, and B. Thierry, "Silicon Nanowires Field Effect Transistors: A Comparative Sensing Performance between Electrical Impedance and Potentiometric Measurement Paradigms," Anal. Chem. 91(19), 12568–12573 (2019).

18 T. Rim, M. Meyyappan, and C.-K. Baek, "Optimized operation of silicon nanowire field effect transistor sensors," Nanotechnology 25(50), 505501 (2014).

19 K. Georgakopoulou, A. Birbas, and C. Spathis, "Modeling of fluctuation processes on the biochemically sensorial surface of silicon nanowire field-effect transistors," Journal of Applied Physics 117(10), 104505 (2015).

20 B.K. Sarker, R. Shrestha, K.M. Singh, J. Lombardi, R. An, A. Islam, and L.F. Drummy, "Label-Free Neuropeptide Detection beyond the Debye Length Limit," ACS Nano 17(21), 20968–20978 (2023).

21 Q. Deng, S. Lv, Z. Li, K. Tan, X. Liang, and S. Shen, "The impact of flexoelectricity on materials, devices, and physics," Journal of Applied Physics 128(8), 080902 (2020).

22 D. Guo, P. Guo, L. Ren, Y. Yao, W. Wang, M. Jia, Y. Wang, L. Wang, Z.L. Wang, and J. Zhai, "Silicon flexoelectronic transistors," Science Advances 9(10), eadd3310 (2023).

23 L. Wang, S. Liu, X. Feng, C. Zhang, L. Zhu, J. Zhai, Y. Qin, and Z.L. Wang, "Flexoelectronics of centrosymmetric semiconductors," Nat. Nanotechnol. 15(8), 661–667 (2020).

24 J. Narvaez, F. Vasquez-Sancho, and G. Catalan, "Enhanced flexoelectric-like response in oxide semiconductors," Nature 538(7624), 219–221 (2016).

25 S.R. Kwon, W.B. Huang, S.J. Zhang, F.G. Yuan, and X.N. Jiang, "Flexoelectric sensing using a multilayered barium strontium titanate structure," Smart Mater. Struct. 22(11), 115017 (2013).

26 X. Liang, S. Hu, and S. Shen, "Effects of surface and flexoelectricity on a piezoelectric nanobeam," Smart Mater. Struct. 23(3), 035020 (2014).

27 J.-L. Fraikin, M.V. Requa, and A.N. Cleland, "Probing the Debye Layer: Capacitance and Potential of Zero Charge Measured Using a Debye-Layer Transistor," Phys. Rev. Lett. 102(15), 156601 (2009).

28 A. Liu, Ph.D. Thesis, Boston University, 2022.

29 B. Lu, M.J. Holst, J.A. McCammon, and Y.C. Zhou, "Poisson-Nernst-Planck Equations for Simulating Biomolecular Diffusion-Reaction Processes I: Finite Element Solutions," J Comput Phys 229(19), 6979–6994 (2010).

30 M. Imboden, and P. Mohanty, "Dissipation in nanoelectromechanical systems," Physics Reports 534(3), 89–146 (2014).

## Acknowledgements

The authors acknowledge support from FemtoDx Inc., Boston University, and the Boston University Photonics Center, where part of this work was performed. This work was also carried out in part at the Harvard University Center for Nanoscale Systems (CNS), a member of the National Nanotechnology Coordinated Infrastructure (NNCI). The NNCI is supported by the National Science Foundation under NSF Award No. ECCS-2025158.

## Ethics declarations
### Competing interests

The authors declare that they have no competing interests.

# Supplementary Information

# Radio Frequency Field-Induced Enhancement of Detection Sensitivity in Silicon Nanowire Sensors

## Author Information

### Affiliations

[1] **Division of Materials Science and Engineering, Boston University, Brookline, Massachusetts 02446, USA**

[2] **Department of Physics, Boston University, 590 Commonwealth Avenue, Boston, Massachusetts 02215, USA**

[3] **FemtoDx, Inc., 8484 Wilshire Blvd, Suite 630, Beverly Hills, CA, 90211, USA**

Ang Liu,[1,3] Jingsong Shang,[2] Jiangang J. Du,[3] Shyamsunder Erramilli,[1,2,3] Pritiraj Mohanty[1,2,3]

### Corresponding author

Correspondence to: Ang Liu, E-Mail: aliu12@bu.edu, Pritiraj Mohanty, E-Mail: mohanty@buphy.bu.edu

This file includes:

Supplementary Figures S1, S2

Supplementary Tables S1 – S4

Supplementary Equations S1 – S5

**Supplementary Figure S1: Focused ion beam preparation of silicon nanowire TEM specimens.**

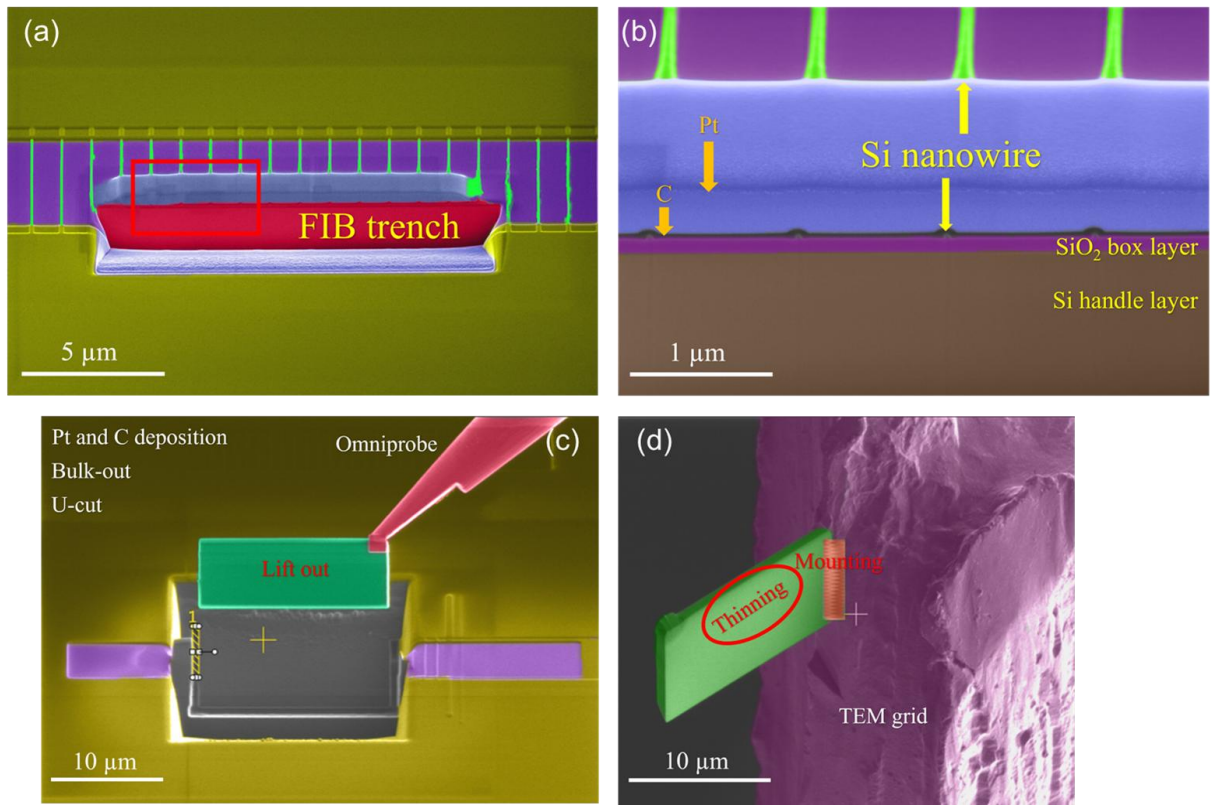

False-colored SEM images illustrating sequential stages of silicon nanowire TEM specimen preparation using the focused ion beam (FIB) technique. The process includes deposition of protective platinum (Pt) and carbon (C) layers, milling to sculpt the sample, and use of an OmniProbe micromechanical probe for precise extraction and transfer of the TEM lamella onto a grid. Subsequent thinning and cleaning steps yield a lamella suitable for high-resolution SEM and TEM imaging, enabling detailed observation of silicon nanowire cross-sections and their interfaces with the silicon dioxide (BOX) substrate. (**a**) FIB trench indicating the region of material removal for TEM sample preparation. (**b**) Tilted SEM image providing a cross-sectional view of silicon nanowires. (**c**) Extraction of a TEM lamella using an OmniProbe following a U-cut. (**d**) TEM lamella mounted on a grid prior to thinning for TEM analysis.

**Supplementary Figure S2: Surface morphology and surface potential evolution during nanowire functionalization.**

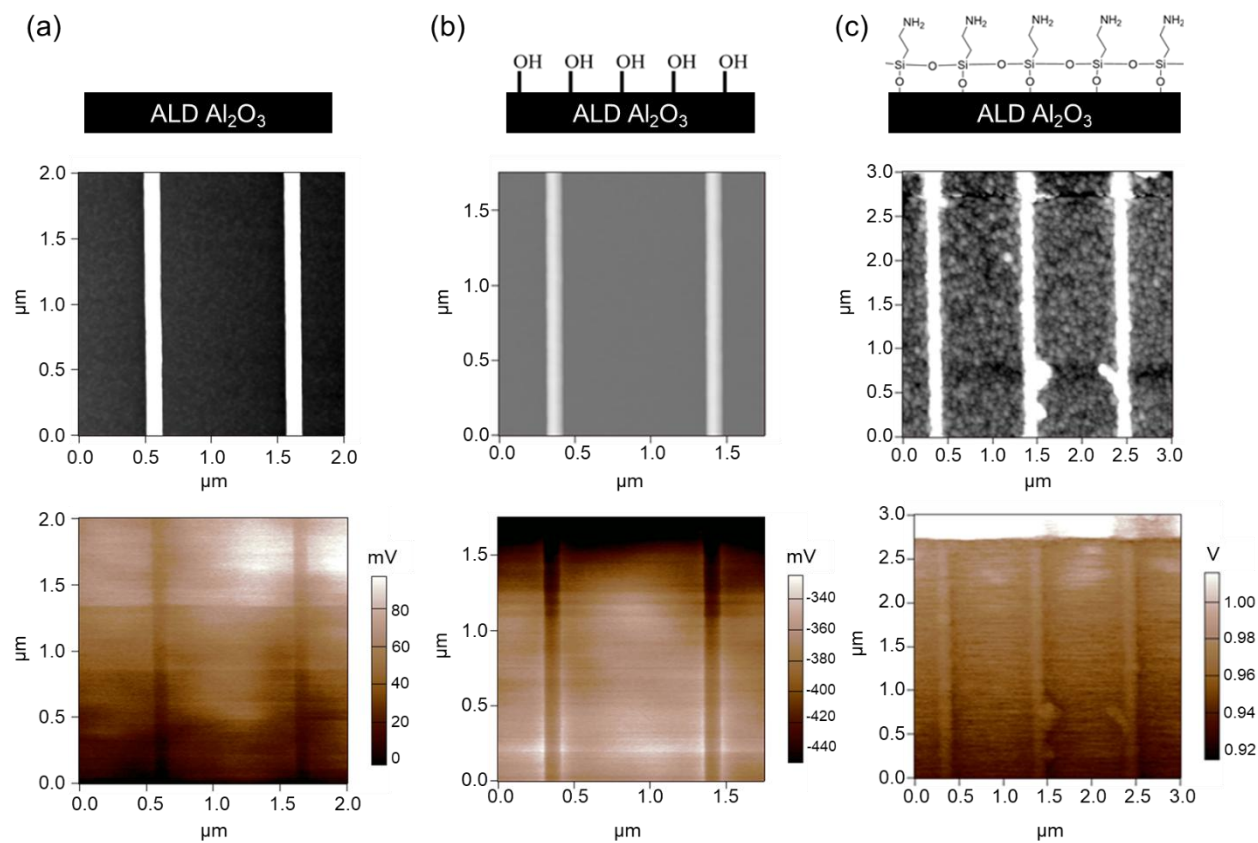

AFM height profiles and Kelvin probe force microscopy (KPFM) surface potential measurements illustrating the surface morphology and potential evolution of Si nanowires on ALD-deposited alumina dielectric during functionalization. Surface morphology and corresponding potential changes are shown following hydroxylation and silanization, processes that are critical for tailoring device properties for sensitive and selective detection of the biomarker C-reactive protein (CRP). Surface characterization of the ALD alumina dielectric at different functionalization stages is presented. (**a**) Pristine ALD surface exhibiting uniform topography and near-neutral surface potential. (**b**) Hydroxylated surface showing minimal topographical change and a slight negative shift in surface potential. (**c**) Silanized surface exhibiting increased roughness due to the presence of primary amine groups and a positive surface potential, indicating readiness for EDC/sulfo-NHS crosslinking with anti-CRP. In each column, the top panel shows the surface functionalization schematic, the middle grayscale images correspond to AFM height profiles, and the bottom images display KPFM surface potential measurements.

**Supplementary Table S1: Surface morphology and surface potential evolution during nanowire functionalization.**

|  | Sample | Central Frequency (MHz) | FWHM (MHz) | Quality Factor | Peak Height (µS) |
|---|---|---|---|---|---|
| 1st Peak | CRP | 10.52 | 2.00 | 5.25 | 73.3 |
|  | PBS | 10.55 | 2.30 | 4.57 | 21.4 |
|  | Dry | 10.96 | 2.13 | 5.12 | 21.5 |
| 1st Trough | CRP | 39.44 | 2.47 | 41.93 | -113 |
|  | PBS | 38.08 | 2.86 | 24.85 | -28.6 |
|  | Dry | 38.28 | 1.70 | 29.31 | -14.1 |
| 2nd Peak | CRP | 64.18 | 3.77 | 50.89 | 132.6 |
|  | PBS | 64.05 | 1.48 | 20.13 | 14.5 |
|  | Dry | 64.52 | 2.15 | 35.12 | 23.3 |
| 2nd Trough | CRP | 124.61 | 3.35 | 103.95 | -263.0 |
|  | PBS | 125.00 | 1.20 | 37.31 | -33.8 |
|  | Dry | 124.93 | 1.72 | 72.63 | -46.6 |
| 3rd Peak | CRP | 191.00 | 4.08 | 46.81 | 12.9 |
|  | PBS | - | - | - | - |
|  | Dry | 186.34 | 3.05 | 61.10 | 16.62 |
| 3rd Trough | CRP | 489.00 | 9.44 | 51.80 | -116 |
|  | PBS | 487.10 | 8.50 | 57.31 | -19.3 |
|  | Dry | 487.60 | 12.50 | 39.01 | -27.4 |

Peak analysis for CRP, wet control (PBS), and dry control groups. This table summarizes differences in peak responses observed in a radiofrequency (RF)-modulated silicon nanowire field-effect transistor (SiNW FET) biosensing platform. The data demonstrate sensor sensitivity and selectivity by comparing conductance variations as a function of RF frequency across sample types. The analysis is supported by Figures 2 and 3, which show CRP concentration-dependent RF spectra and corresponding peak area and baseline responses, validating specific and non-specific binding during CRP detection. Columns represent key signal characteristics: (**1**) Central/Peak Frequency - the resonant frequency at which the response peak occurs; (**2**) FWHM ($\Delta f$) - full width at half maximum, describing the bandwidth over which the response is at least half of its maximum; (**3**) Quality Factor - the ratio of central frequency to FWHM, indicating resonator selectivity and damping; and (**4**) Peak Height - the maximum response amplitude, correlated with analyte concentration.

**Supplementary Table S2: Measured and theoretical flexural mode frequencies**

| Flexural Mode # | Measured Frequency (MHz) | Theoretical Frequency (MHz) | Deviation |
|---|---|---|---|
| 1 | 3.52 | Base Frequency | NA |
|  | 6.76 |  |  |
| 2 | 10.98 | $3.52 * \frac{(2*2+1)^2}{(2*1+1)^2} = 9.78$ | 12% |
| 3 | 21.73 | $3.52 * \frac{(2*3+1)^2}{(2*1+1)^2} = 19.16$ | 13% |
| 4 | 35.56 | $3.52 * \frac{(2*4+1)^2}{(2*1+1)^2} = 31.68$ | 12% |
| 5 | 40.32 | $3.52 * \frac{(2*5+1)^2}{(2*1+1)^2} = 47.32$ | 14% |
|  | 60.42 |  |  |
| 6 | 68.47 | $3.52 * \frac{(2*6+1)^2}{(2*1+1)^2} = 66.10$ | 3% |
|  | 76 |  |  |
| 7 | 85.35 | $3.52 * \frac{(2*7+1)^2}{(2*1+1)^2} = 88.00$ | 3% |
| 8 | 102.61 | $3.52 * \frac{(2*8+1)^2}{(2*1+1)^2} = 113.03$ | 9% |
|  | 124.61 |  |  |
| 9 | 133.56 | $3.52 * \frac{(2*9+1)^2}{(2*1+1)^2} = 141.19$ | 5% |
| 10 | 165.2 | $3.52 * \frac{(2*10+1)^2}{(2*1+1)^2} = 172.47$ | 4% |
|  | 191 |  |  |
| 11 | 206.86 | $3.52 * \frac{(2*11+1)^2}{(2*1+1)^2} = 206.89$ | 0% |
| 12 | 244.66 | $3.52 * \frac{(2*x+1)^2}{(2*1+1)^2} = 244.44$ | 0% |
|  | 407.19 |  |  |
| 16 | 416.06 | $3.52 * \frac{(2*16+1)^2}{(2*1+1)^2} = 425.91$ | 2% |
|  | 445.15 |  |  |
| 17 | 489 | $3.52 * \frac{(2*17+1)^2}{(2*1+1)^2} = 479.10$ | 2% |
| 18 | 526.86 | $3.52 * \frac{(2*18+1)^2}{(2*1+1)^2} = 535.42$ | 1% |
|  | 532.00 |  |  |
| 23 | 877.8 | $3.52 * \frac{(2*23+1)^2}{(2*1+1)^2} = 863.94$ | 1% |

Flexural mode analysis with a base frequency of 3.52 MHz. This table compares measured and theoretical resonance frequencies across multiple mode numbers. Deviation percentages quantify the agreement between experiment and theory, where theoretical frequencies are calculated using the relation $\frac{(2*n+1)^2}{(2*1+1)^2}$. This comparison provides insight into the flexoelectric resonance behavior of the system and demonstrates the reliability of the flexural mode model under the given conditions.

**Supplementary Table S3: Flexoelectric coefficients of silicon (Si). Only symmetry-independent coefficients are reported.**

| Flexoelectric coefficients | Centrosymmetric semiconductor, Si |
|---|---|
| $\mu_{1111}$ (nC/m) | 1.3 |
| $\mu_{1122}$ (nC/m) | 0.4 |
| $\mu_{1221}$ (nC/m) | 0.4 |
| $\mu_{1133}$ (nC/m) | - |
| $\mu_{2332}$ (nC/m) | - |
| $\mu_{3113}$ (nC/m) | - |
| $\mu_{3311}$ (nC/m) | - |
| $\mu_{3333}$ (nC/m) | - |

**Supplementary Table S4: Material parameters of silicon (Si) used in COMSOL Multiphysics simulations.**

| Parameters | Centrosymmetric semiconductor, Si |
|---|---|
| $\rho \ [kg \cdot m^{-3}]$ | 2329 |
| $E \ [Pa]$ | $1.7 * 10^{11}$ |
| $\nu$ | 0.28 |
| $\varepsilon$ | 11.9 |
| Doping $[cm^{-3}]$ | $5 * 10^{16}$ |
| $n_i \ [cm^{-3}]$ | $1.02 * 10^{10}$ |
| $\mu_n \ [cm^2/(V \cdot s)]$ | 1450 |
| $\mu_p \ [cm^2/(V \cdot s)]$ | 500 |
| $E_a \ [eV]$ | 4.05 |
| $E_g \ [eV]$ | 1.12 |
| $\Phi_B \ [eV]$ | 0.51(n)/0.55(p) |
| $\tau_n \ [ns]$ | 100 |
| $\tau_p \ [ns]$ | 100 |

**Supplementary Equation S1: Polarization vector.**

The polarization vector, P, describes the electric dipole moment per unit volume within a material. In flexoelectric systems, polarization is influenced by mechanical deformation in the form of strain gradients and can be expressed as:

$$P_i = P_i^{intrinsic} + P_i^{flexo},$$

where $P_i^{intrinsic}$ represents polarization arising from intrinsic material properties (e.g., spontaneous polarization in ferroelectrics), and $P_i^{flexo}$ denotes the component induced by strain gradients. For centrosymmetric materials, where $P_i^{intrinsic} = 0$, the polarization is governed entirely by the flexoelectric contribution.

**Supplementary Equation S2: Flexoelectric polarization.**

The flexoelectric effect describes a linear coupling between strain gradients and polarization, expressed as:

$$P_i^{flexo} = f_{ijkl} \frac{\partial \varepsilon_{jk}}{\partial x_l},$$

where $P_i^{flexo}$ is the polarization induced in the i-th direction, $f_{ijkl}$ is the fourth-rank flexoelectric tensor characterizing the electromechanical coupling, $\varepsilon_{jk}$ is the strain tensor, and $\frac{\partial \varepsilon_{jk}}{\partial x_l}$ denotes the spatial gradient of strain. This induced polarization becomes particularly significant at the nanoscale, where large strain gradients are present.

**Supplementary Equation S3: Strain tensor.**

The strain tensor, $\varepsilon_{jk}$, measures the local deformation within a material and is derived from the displacement field $\mathbf{u} = (u_x, u_y, u_z)$:

$$\varepsilon_{jk} = \frac{1}{2}\left(\frac{\partial u_j}{\partial x_k} + \frac{\partial u_k}{\partial x_j}\right)$$

where:

- $u_j$ and $u_k$ are displacement components along the *j*- and *k*-axes, respectively.
- $x_j$ and $x_k$ are spatial coordinates within the material.

This symmetric tensor captures the material's deformation due to mechanical forces.

**Supplementary Equation S4: Strain gradient.**

The strain gradient, $\frac{\partial \varepsilon_{jk}}{\partial x_l}$, quantifies the spatial variation of strain within a material and serves as a key driver of the flexoelectric effect. Substituting the definition of the strain tensor, the strain gradient can be expressed as:

$$\frac{\partial \varepsilon_{jk}}{\partial x_l} = \frac{1}{2}\left(\frac{\partial^2 u_j}{\partial x_k \partial x_l} + \frac{\partial^2 u_k}{\partial x_j \partial x_l}\right)$$

This higher-order spatial derivative is sensitive to nanoscale variations, making it a pivotal quantity in flexoelectric behavior.

**Supplementary Equation S5: Final equation for induced polarization.**

Substituting the strain gradient into the polarization equation, the flexoelectric-induced polarization becomes:

$$P_i^{flexo} = \frac{1}{2} f_{ijkl}\left(\frac{\partial^2 u_j}{\partial x_k \partial x_l} + \frac{\partial^2 u_k}{\partial x_j \partial x_l}\right)$$